\documentclass[twocolumn,aps,showpacs,superscriptaddress]{revtex4}
\usepackage{graphicx}

\newcommand{\srt}{$\sqrt{s_{_{\rm NN}}}$}

\newcommand{\pt}{$p_T$ }

\newcommand{\GeVc}{GeV/$c$ }

\begin{document}
% A useful Journal macro
\def\Journal#1#2#3#4{{#1} {\bf #2}, #3 (#4)}

% Some useful journal names
\def\NCA{\em Nuovo Cimento}
\def\NIM{\em Nucl. Instr. Meth.}
\def\NIMA{{\em Nucl. Instr. Meth.} A}
\def\NPB{{\em Nucl. Phys.} B}
\def\NPA{{\em Nucl. Phys.} A}
\def\PLB{{\em Phys. Lett.} B}
\def\PRL{{\em Phys. Rev. Lett.}}
\def\PRC{{\em Phys. Rev.} C}
\def\PRD{{\em Phys. Rev.} D}
\def\ZPC{{\em Z. Phys.} C}
\def\JPG{{\em J. Phys.} G}
\def\EPJ{{\em Eur. Phys. J.} C}
\def\RPP{{\em Rep. Prog. Phys.}}
\def\IPA{{\em Int. J. Mod. Phys.} A}
\def\IPE{{\em Int. J. Mod. Phys.} E}

\preprint{}
\title{Open charm yields in d+Au collisions at $\sqrt{s_{_{NN}}}=200$ GeV}

\affiliation{Argonne National Laboratory, Argonne, Illinois 60439}
\affiliation{University of Bern, 3012 Bern, Switzerland}
\affiliation{University of Birmingham, Birmingham, United Kingdom}
\affiliation{Brookhaven National Laboratory, Upton, New York
11973} \affiliation{California Institute of Technology, Pasadena,
California 91125} \affiliation{University of California, Berkeley,
California 94720} \affiliation{University of California, Davis,
California 95616} \affiliation{University of California, Los
Angeles, California 90095} \affiliation{Carnegie Mellon
University, Pittsburgh, Pennsylvania 15213} \affiliation{Creighton
University, Omaha, Nebraska 68178} \affiliation{Nuclear Physics
Institute AS CR, 250 68 \v{R}e\v{z}/Prague, Czech Republic}
\affiliation{Laboratory for High Energy (JINR), Dubna, Russia}
\affiliation{Particle Physics Laboratory (JINR), Dubna, Russia}
\affiliation{University of Frankfurt, Frankfurt, Germany}
\affiliation{Institute  of Physics, Bhubaneswar 751005, India}
\affiliation{Indian Institute of Technology, Mumbai, India}
\affiliation{Indiana University, Bloomington, Indiana 47408}
\affiliation{Institut de Recherches Subatomiques, Strasbourg,
France} \affiliation{University of Jammu, Jammu 180001, India}
\affiliation{Kent State University, Kent, Ohio 44242}
\affiliation{Lawrence Berkeley National Laboratory, Berkeley,
California 94720} \affiliation{Massachusetts Institute of
Technology, Cambridge, MA 02139-4307}
\affiliation{Max-Planck-Institut f\"ur Physik, Munich, Germany}
\affiliation{Michigan State University, East Lansing, Michigan
48824} \affiliation{Moscow Engineering Physics Institute, Moscow
Russia} \affiliation{City College of New York, New York City, New
York 10031} \affiliation{NIKHEF, Amsterdam, The Netherlands}
\affiliation{Ohio State University, Columbus, Ohio 43210}
\affiliation{Panjab University, Chandigarh 160014, India}
\affiliation{Pennsylvania State University, University Park,
Pennsylvania 16802} \affiliation{Institute of High Energy Physics,
Protvino, Russia} \affiliation{Purdue University, West Lafayette,
Indiana 47907} \affiliation{University of Rajasthan, Jaipur
302004, India} \affiliation{Rice University, Houston, Texas 77251}
\affiliation{Universidade de Sao Paulo, Sao Paulo, Brazil}
\affiliation{University of Science \& Technology of China, Anhui
230027, China} \affiliation{Shanghai Institute of Applied Physics,
Shanghai 201800, China} \affiliation{SUBATECH, Nantes, France}
\affiliation{Texas A\&M University, College Station, Texas 77843}
\affiliation{University of Texas, Austin, Texas 78712}
\affiliation{Tsinghua University, Beijing 100084, China}
\affiliation{Valparaiso University, Valparaiso, Indiana 46383}
\affiliation{Variable Energy Cyclotron Centre, Kolkata 700064,
India} \affiliation{Warsaw University of Technology, Warsaw,
Poland} \affiliation{University of Washington, Seattle, Washington
98195} \affiliation{Wayne State University, Detroit, Michigan
48201} \affiliation{Institute of Particle Physics, CCNU (HZNU),
Wuhan 430079, China} \affiliation{Yale University, New Haven,
Connecticut 06520} \affiliation{University of Zagreb, Zagreb,
HR-10002, Croatia}

\author{J.~Adams}\affiliation{University of Birmingham, Birmingham, United Kingdom}
\author{M.M.~Aggarwal}\affiliation{Panjab University, Chandigarh 160014, India}
\author{Z.~Ahammed}\affiliation{Variable Energy Cyclotron Centre, Kolkata 700064, India}
\author{J.~Amonett}\affiliation{Kent State University, Kent, Ohio 44242}
\author{B.D.~Anderson}\affiliation{Kent State University, Kent, Ohio 44242}
\author{D.~Arkhipkin}\affiliation{Particle Physics Laboratory (JINR), Dubna, Russia}
\author{G.S.~Averichev}\affiliation{Laboratory for High Energy (JINR), Dubna, Russia}
\author{S.K.~Badyal}\affiliation{University of Jammu, Jammu 180001, India}
\author{Y.~Bai}\affiliation{NIKHEF, Amsterdam, The Netherlands}
\author{J.~Balewski}\affiliation{Indiana University, Bloomington, Indiana 47408}
% ...

\collaboration{STAR Collaboration}\noaffiliation
%\collaboration{STAR
%Collaboration}\homepage{www.star.bnl.gov}\noaffiliation

\date{\today}% It is always \today, today,
\begin{abstract}

Mid-rapidity open charm spectra from direct reconstruction of
$D^{0}$($\overline{D^0}$)$\rightarrow K^{\mp}\pi^{\pm}$ in d+Au
collisions and indirect electron/positron measurements via charm
semileptonic decays in p+p and d+Au collisions at \srt = 200 GeV
are reported. The $D^{0}$($\overline{D^0}$) spectrum covers a
transverse momentum ($p_T$) range of 0.1 $<p_T<$ 3 \GeVc whereas
the electron spectra cover a range of 1 $<p_T<$ 4 GeV/$c$. The
electron spectra show approximate binary collision scaling between
p+p and d+Au collisions. From these two independent analyses, the
differential cross section per nucleon-nucleon binary interaction
at mid-rapidity for open charm production from d+Au collisions at
RHIC is $d\sigma^{NN}_{c\bar{c}}/dy$=0.30$\pm$0.04
(stat.)$\pm$0.09(syst.) mb. The results are compared to
theoretical calculations. Implications for charmoniumm results in
A+A collisions are discussed.
\end{abstract}
\maketitle

%--===========================================================
%\section{Introductions}

Hadrons with heavy flavor are unique tools for studying the strong
interaction described by Quantum Chromodynamics (QCD). Due to the
large mass of the charm quark ($\sim$ 1.5 GeV/$c^{2}$), charm
quark production can be evaluated by perturbative QCD (pQCD) even
at low momentum through the introduction of additional scales
related to the charm quark mass~\cite{charm1,charm2}. Therefore,
theoretical calculation of charm hadron total cross section
integrated over momentum space is expected to be less affected by
non-perturbative soft processes and hadronization~\cite{pQCDpi0}.
Systematic studies of charm production in p+p and p+nucleus
collisions have been proposed as a sensitive way to measure the
parton distribution function in nucleons, and nuclear shadowing
effects~\cite{lin96}. At RHIC energies, heavy quark energy
loss~\cite{dokshitzer01}, charm quark
coalescence~\cite{loic,pbm,rafelski,mclerran}, possible $J/\psi$
suppression~\cite{matsui}, and charm flow~\cite{xu2} have been
proposed as important tools in studying the properties of matter
created in heavy ion collisions.

Identification of charmed hadrons is difficult due to their short
lifetime ($c\tau(D^{0})=124$ $\mu$m), low production rates, and
large combinatorial background. Most measurements of the total
charm cross section in hadron-hadron collisions have been
performed at low center-of-mass energies ( $\lesssim$ 40 GeV) in
fixed target experiments~\cite{tavernier,fermic}. At
$\sqrt{s}\sim52-63$ GeV, the available measurements are not
conclusive due to inconsistencies between different
measurements~\cite{tavernier,isrc}. The measurements at higher
energy colliders have been at high $p_{T}$ only~\cite{CDFII} , or
have included large uncertainties~\cite{phenixe,ua2}. Theoretical
predictions for the RHIC energy region differ
significantly~\cite{vogt02, dipole}. Therefore, precise
measurements of charm cross sections in p+p and d+Au collisions in
this energy region are crucial.  In this paper, we report first
results on open charm cross sections at \srt = 200 GeV from direct
charmed hadron $D^0$($\overline{D^0}$) reconstruction in d+Au
collisions and from charm semileptonic decay in both p+p and d+Au
collisions. These measurements are complementary, providing
important experimental cross-checks.

%--===========================================================
%\section{Analysis methods and results}
%\underline{\bf D$^0$ Reconstruction}
The data used in $D^0$ direct reconstruction and charm
semileptonic decay analysis were taken during the 2003 RHIC run in
d+Au and p+p collisions at \srt = 200 GeV with the Solenoidal
Tracker at RHIC (STAR). A minimum bias d+Au collision trigger was
defined by requiring at least one spectator neutron in the
outgoing Au beam direction depositing energy in a Zero Degree
Calorimeter (ZDC). Detailed descriptions of the trigger and
centrality definition in d+Au collisions have been presented in a
previous publication~\cite{dAuhighpt}. A total of 15.7 million
minimum bias triggered d+Au collision events were used in the
$D^{0}$ analysis. The data samples used in the electron analysis
in d+Au and p+p collisions were described in Ref.~\cite{startof1}.
The integrated luminosity is about 40 $\mu$b$^{-1}$ for d+Au
collisions and 30 nb$^{-1}$ for p+p collisions.
%The normalization
%uncertainties of total cross sections in p+p and d+Au collisions
%are 14\% and 16\% respectively.

\begin{figure}[h]
\centerline{\includegraphics[width=0.5\textwidth] {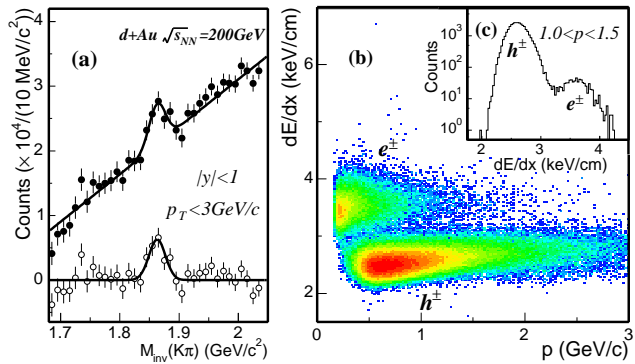}}
 \caption[]{(a) Invariant mass distributions of kaon-pion pairs from
 d+Au collisions. The solid circles depict the signal after mixed-event
 background subtraction, the open circles after subtraction
 of the residual background using a linear
 parametrization.
 %The fit range is in 1.68 $<M_{inv}<$ 2.05 GeV/$c^2$.
 (b) $dE/dx$ in the TPC
 vs. particle momentum ($p$) with a TOF cut of $|1/\beta - 1| \le 0.03$.
 Insert: projection on the $dE/dx$ axis for particle momenta 1$<p<$1.5 GeV/$c$.}
 \label{fig1pid}
\end{figure}

The primary tracking device of the STAR detector is the Time
Projection Chamber (TPC)~\cite{tpcnim}. It was used to reconstruct
the decay of $D^0\rightarrow K^-\pi^+$ ($\overline{D^0}\rightarrow
K^+\pi^-$) which has a branching ratio of 3.83\%. In what follows,
we imply $(D^0+\overline{D^0})/2$ when using the term $D^0$ unless
otherwise specified. The exact $D^{0}$ decay topology cannot be
resolved due to insufficient track projection resolution close to
the collision vertex. The invariant mass spectrum of $D^0$ mesons
was obtained by pairing each oppositely charged kaon and pion
candidate in the same event. The kaon and pion tracks were
identified through ionization energy loss ($dE/dx$) in the TPC
wherever the identification is possible. Candidate tracks were
selected having momenta $p$ ($p_{T}$) $>0.3$ ($0.2$) GeV/$c$ and
pseudorapidity $|\eta|<1$.
%The $D^{0}$ and $\overline{D^0}$ spectra were added
%together in this analysis in order to increase statistics.
The $D^0$ signal with $p_T<3$ GeV/$c$ and $|y|<1$ after
mixed-event background subtraction~\cite{kstarprc} is shown in the
left panel of Fig.~\ref{fig1pid}. The signal-to-background ratio
($S/N$) is about 1/600, and the figure of merit ($S/\sqrt{N}$) is
about 6.
%There remains residual background not described by the
%mixed-event spectrum. HIJING simulations~\cite{hijing} show that
%di-hadron correlations from jets can significantly affect the line
%shape of the background spectrum.
%In addition to the $D^{0}$ signal, there remains residual
%background, which may be due to di-hadron correlation from jets,
%cannot be described by the mixed-event spectrum.
This distribution was fit to a Gaussian plus a linear function to
account for the residual background not described by the
mixed-event spectrum~\cite{kstarprc}. The open symbols in the left
panel of Fig.~\ref{fig1pid} depict the $D^0$ signal after the
two-step background subtraction. HIJING simulations~\cite{hijing}
have shown that di-hadron correlations from jets can affect the
line-shape of the background spectrum since the shape (slope
versus mass) from this contribution is different from that of
random pairs. To estimate the uncertainty in the subtraction of
the residual background, different normalizations, slopes and fit
ranges were tried. The resulting uncertainty in the $D^0$ yield is
estimated to be 15\%.
%\vspace{-0.5cm}
Within statistical uncertainties, the yields of $D^0$ and
$\overline{D^0}$ are equal. The $D^0\rightarrow K^{-}\pi^{+}$
signal could be mis-identified as a $\overline{D^0}\rightarrow
K^{+}\pi^{-}$ and vice versa when both of its daughters are beyond
particle identification in the TPC. This misidentification results
in double counting which was corrected for in the $D^0$ yields
through a Monte Carlo simulation.

%\underline{\bf Electron measurements}
Another detector used in this analysis was a prototype
time-of-flight system (TOFr)~\cite{tofnim} based on multi-gap
resistive plate chamber technology. It covers an azimuthal angle
$\Delta\phi\simeq\pi/30$, and $-1<\eta<0$. In addition to its
hadron identification capability~\cite{startof1}, it allows
electrons/positrons to be identified at low momentum ($p_{T}<3$
GeV/$c$) by using a combination of velocity information ($\beta$)
from TOFr and $dE/dx$ measured in the TPC. The right panel of
Fig.~\ref{fig1pid} demonstrates the clean separation of electrons
from hadrons using their $dE/dx$ in the TPC after applying a TOFr
cut of $|1/\beta - 1| \le 0.03$. This cut eliminated the hadrons
crossing the electron $dE/dx$ band. Electrons/positrons were
required to originate from the collision vertex.  Hadron
contamination was evaluated to be about $10-15$\% in a selection
optimized for purity and statistics.  At higher $p_{T}$ ($2-4$
GeV/$c$), electrons could be identified directly in the TPC since
hadrons have lower $dE/dx$ due to the relativistic rise of the
$dE/dx$ for electrons.
%, which can be done in the minimum bias triggered data sets to get more statistics.
Positrons are more difficult to identify using $dE/dx$ alone
because of the large background from the deuteron band.
%Only those
%electrons with $dE/dx$ above the expected mean values were
%selected to reduce hadron contamination.
The hadron contamination in this case was found to be $\lesssim$
5\% at $p_{T}\simeq2$ GeV/$c$ and to increase to $\sim$ 30\% at
$p_{T}\simeq3-4$ GeV/$c$. This was corrected for in the final
spectra.
%--=======================================================
%\vspace{-0.5cm}
\begin{figure}[h] \centerline{\includegraphics[width=0.475\textwidth]
{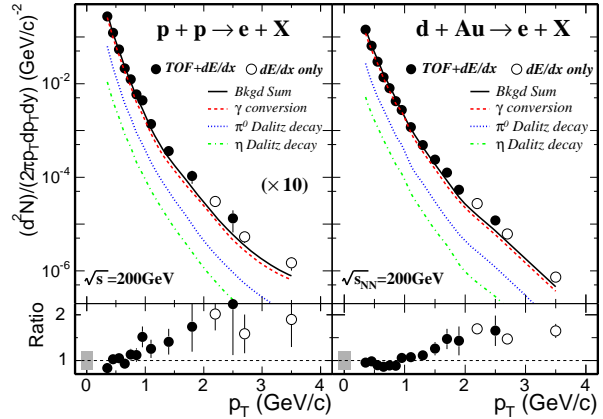}} \caption[]{Upper panels: Electron distributions
from p+p (left) and d+Au (right) collisions. Solid and open
symbols depict electrons/positrons (($e^{+}$+$e^{-}$)/2)
identified via a combination of Time-of-Flight (TOF) and $dE/dx$,
and electrons ($e^{-}$) identified via $dE/dx$ alone. The total
photonic backgrounds are shown as solid lines. Dashed lines depict
the various contributing sources. The fractions were derived from
simulations.
%, showing more than 95\% of photonic background sources
%were measured
Bottom panels: the ratio of inclusive electrons to the total
backgrounds. The gray band represents the systematic uncertainty
in each panel.} \label{fig2einc}
\end{figure}
%--=====================================================
%The correction for hadron contamination was evaluated to be $5-30\%$, depending on $p_{T}$.
Detector acceptance and efficiency corrections were determined
from detailed simulations~\cite{startof1}. Total inclusive
electron spectra from 200 GeV p+p and d+Au collisions are shown in
Fig.~\ref{fig2einc}.

Gamma conversions $\gamma\rightarrow e^{+}e^{-}$ and
$\pi^0\rightarrow\gamma e^{+}e^{-}$ Dalitz decays are the dominant
photonic sources of electron background.
%The $e^{+}e^{-}$ pairs
%from these processes are present mainly at small pair invariant
%mass and/or small opening angle.
%Due to the large coverage of the TPC, the efficiency of finding
%$e^{+}e^{-}$ pairs is very high for such processes.
To measure the background photonic electron spectra, the invariant
mass and opening angle of the $e^+e^-$ pairs were constructed from
an electron (positron) in TOFr and every other positron (electron)
candidate reconstructed in the TPC~\cite{johnson}. A secondary
vertex at the conversion point was not required. Simulations with
both HIJING \cite{hijing} and PYTHIA~\cite{pythia1} with full
detector description in GEANT yielded $\sim$ 60\% efficiency for
electrons with $p_{T}>1$
GeV/$c$ from such background processes. %We found the \pt
%ependence negligible. This value was used to correct for the
%photonic electron spectra in the data.
More than 95\% of the electrons from sources other than
heavy-flavor semileptonic decays were measured with this method.
The remaining fraction from decays of $\eta, \omega,
\rho, \phi$ and $K$ was determined from simulations. The results
are shown as solid lines
in Fig.~\ref{fig2einc}. % We found that the probability of randomly
%rejecting a signal electron with this method was about 40\% at
%$p_{T}\sim0.5$ GeV/$c$ and less than 10\% at $p_{T}\sim3$ GeV/$c$.
The overall uncertainty of the background is on the order of 20\%
and has been included in the systematic errors.
%of the final electron distributions
Ratios of the inclusive electrons over the total backgrounds are
shown in the bottom panels of Fig.~\ref{fig2einc}. The signal is
clearly in excess of the background above $p_{T}>1$ GeV/$c$.

%--=======================================================
%\vspace{-0.3cm}
\begin{figure}[h] \centerline{\includegraphics[width=0.45\textwidth]
{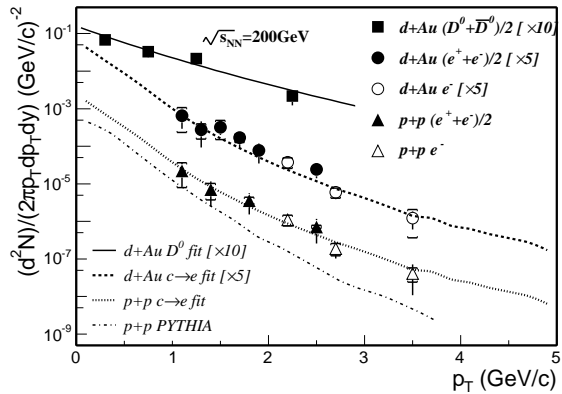}} \caption[]{ Reconstructed $D^0$ (solid squares)
$p_{T}$ distributions from d+Au collisions at $\sqrt{s_{NN}}$ =
200 GeV. Non-photonic electron \pt distributions from p+p
collisions (triangles) and d+Au collisions (circles). Solid and
dashed lines are the fit results from both $D^{0}$ and electron
spectra in d+Au collisions. The dotted line is scaled down by a
factor of $N_{bin}=7.5\pm0.4$ ~\cite{dAuhighpt} from d+Au to p+p
collisions. The dot-dashed line depicts a PYTHIA
calculation~\cite{pythia1}. } \label{fig3de}
\end{figure}

The non-photonic electron spectra were obtained by subtracting the
previously described photonic background from the inclusive
spectra. The results are shown in Fig.~\ref{fig3de}. The $D^0$
invariant yields $d^2N/(2\pi p_Tdp_Tdy)$ as a function of $p_T$
from direct reconstruction are shown in Fig.~\ref{fig3de} as solid
squares. Two different fitting methods were used to extract
$dN/dy$ for the $D^0$ at mid-rapidity. In the first method,
$dN/dy$ was extracted from an exponential fit to the $D^0$
differential yield in transverse mass ($m_T$)~\cite{kstarprc}. In
the second method, a simultaneous fit was applied to both directly
reconstructed $D^0$'s and the background subtracted non-photonic
electron distribution in d+Au collisions. For this fit, it was
assumed that the $D^{0}$ spectrum follows a power law in $p_{T}$
from which an electron spectrum was generated using the particle
composition from~\cite{pdg} and the decay generators in PYTHIA. A
set of parameters for the power law was found at the minimum of
$\chi^{2}$ for the $D^0$ and electron spectra. The results are
shown in Table~\ref{tab1}. The systematic error is dominated by
the uncertainties in the background subtraction, the extrapolation
due to finite $p_{T}$ coverage, and the overall normalization
($\pm14\%$ in p+p and $\pm10\%$ in d+Au
collisions~\cite{dAuhighpt,startof1}).

The yield of $D^{0}$ at mid-rapidity is
$dN/dy=0.028\pm0.004\pm0.008$ and the $\langle p_T \rangle =
1.32\pm0.08$ GeV/$c$ in d+Au collisions. We used the ratio
$R=N_{D^{0}}/N_{c\bar{c}}=0.54\pm0.05$ from $e^{+}e^{-}$ collider
data~\cite{pdg} to convert the $D^{0}$ yield to a total $c\bar{c}$
yield.
%--=======================================================
%\vspace{-0.2cm}
\begin{figure}[ht]
\centerline{\includegraphics[width=0.45\textwidth]{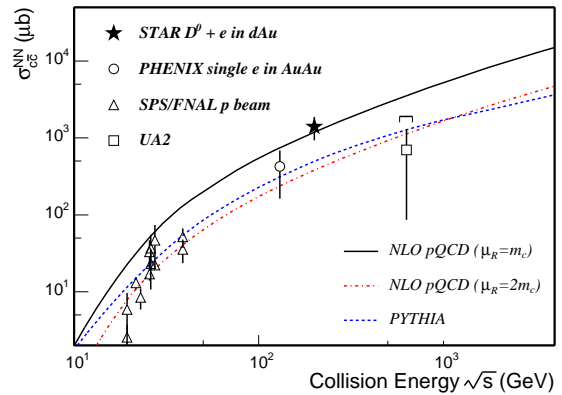}}
\caption[]{Total $c\bar{c}$ cross section per nucleon-nucleon
collision vs.~the collision energy (\srt). The dashed line depicts
a PYTHIA calculation~\cite{pythia1}. The solid and dot-dashed
lines depict two NLO pQCD calculations with MRST HO, $m_{c}$ = 1.2
GeV/$c^{2}$, $\mu_{F}=2m_{c}$, $\mu_{R}=m_{c}$ and $2m_{c}$,
respectively~\cite{vogt02}.} \label{fig4e}
\end{figure}
%--=======================================================
A p+p inelastic scattering cross section of $\sigma^{pp}_{inel}=42$ mb
was used in the calculation and a factor of $f=4.7\pm0.7$, estimated
from simulation \cite{vogt02,pythia1}, was used to convert the
$d\sigma/dy$ at mid-rapidity to the total cross section. The total
charm cross section per nucleon-nucleon interaction for d+Au
collisions at 200 GeV is
$\sigma^{NN}_{c\bar{c}}=dN^{d+Au}_{D^{0}}/dy\times\sigma^{pp}_{inel}
/N^{d+Au}_{bin}\times f/R=1.3\pm0.2\pm0.4$ mb from $D^{0}$ alone and
$1.4\pm0.2\pm0.4$ mb from the combined fit of $D^0$ and electrons. The
nuclear modification factor~\cite{dAuhighpt} was obtained by taking
the ratio of the electron spectra in d+Au and p+p collisions scaled
with the underlying nucleon-nucleon binary collisions. It was measured
to be $1.3\pm0.3\pm0.3$, averaged over $1<p_T<4$ GeV/$c$. This value
is consistent with binary scaling within the measured errors.

\begin{table}[htbp]
\centerline{
\begin{tabular}{c|c|c} \hline
  & $dN(D^{0})$/$dy|_{y=0}$ ($10^{-2}$)  & $d\sigma_{c\bar{c}}^{NN}$/$dy|_{y=0}$ (mb) \\ \hline
  $D^0$         & 2.8$\pm$0.4$\pm$0.8  & 0.29$\pm$0.04$\pm$0.08 \\ \hline
  $D^0$+$e^{\pm}$ & 2.9$\pm$0.4$\pm$0.8 & 0.30$\pm$0.04$\pm$0.09 \\ \hline
%  $D$ +$e^{\pm}$ & 2.7$\pm$0.3$\pm$0.7 & 0.28$\pm$0.03$\pm$0.08 \\ \hline
\end{tabular}}
\caption{$dN/dy$ of $D^{0}$ in d+Au collisions and the corresponding
$d\sigma/dy$ of $c\bar{c}$ pair per nucleon-nucleon collision at \srt
= 200 GeV.} \label{tab1}
\end{table}
%--===========================================================
%\section{Discussions}
The beam energy dependence of the cross section is shown in
Fig.~\ref{fig4e}. Both default PYTHIA~\cite{pythia1} and NLO
pQCD~\cite{vogt02} calculations reasonably describe the results at
lower energies, but underpredict the total charm cross section at
$\sqrt{s_{_{NN}}}=200$ GeV. A NLO pQCD calculation (solid line)
with fragmentation and renormalization scales chosen to be
$\mu_F=2m_{c}$ and $\mu_R=m_{c}$ ($m_{c}=1.2$ GeV$/c^2$)
reproduces our result.
%but overpredict the total charm cross
%sections at other energies. This might indicate the energy
%dependence of scales in those calculations.
The underprediction by PYTHIA of the charm cross section is also
evident in Fig.~\ref{fig3de}
% from comparison of the predicted and measured
%\pt distributions for electrons from charm semileptonic decays.
, the charm decayed electron $p_T$ distribution shown as
dot-dashed line. Furthermore, the slope of the PYTHIA
distributions is much steeper than the measured distribution.
There are also indications that a large charm production cross
section at $\sqrt{s_{_{NN}}}\simeq300$ GeV is essential to explain
available cosmic ray data~\cite{cosmic}.

At RHIC energies, binary scaling of the open charm production is
expected between p+p, p+A and A+A collisions~\cite{lin96}. If
correct, the results of this study suggest a much larger charm
yield in central Au+Au collisions than previously assumed in
statistical thermal models~\cite{pbm,rafelski,mclerran} based on
some pQCD/PYTHIA calculations. This would rule out several
predictions~\cite{pbm,rafelski,mclerran} of charm production not
previously excluded by the upper limit (below binary scaling) set
by $J/\psi$ production in central Au+Au
collisions~\cite{phenixjpsi}. Future heavy ion runs at RHIC with
open charm and $J/\psi$ measurements will enable us to study the
flow and thermalization of charmed particles.

%--===========================================================
%\section{Summary}

In summary, the charm cross section and transverse momentum
distribution for p+p and d+Au collisions at \srt = 200 GeV have
been measured by the STAR collaboration at RHIC. Independent
measurements of the reconstructed $D^0$ and single electrons from
charm semileptonic decay are consistent. The total cross section
at this energy was compared to theoretical calculations. The
result has important consequences for charm quark coalescence in
Au+Au collisions at RHIC.

%--=======================================================

We are grateful for many fruitful discussions with D. Kharzeev,
and R. Vogt. We thank the RHIC Operations Group and RCF at BNL,
and the NERSC Center at LBNL for their support. This work was
supported in part by the HENP Divisions of the Office of Science
of the U.S. DOE; the U.S. NSF; the BMBF of Germany; IN2P3, RA,
RPL, and EMN of France; EPSRC of the United Kingdom; FAPESP of
Brazil; the Russian Ministry of Science and Technology; the
Ministry of Education and the NNSFC of China; Grant Agency of the
Czech Republic, FOM and UU of the Netherlands, DAE, DST, and CSIR
of the Government of India; Swiss NSF; and the Polish State
Committee for Scientific Research.

%--=======================================================


\begin{thebibliography}{999}
% ---
\bibitem{charm1} P.L. McGaughey {\it et al.},
\Journal{\IPA}{10}{2999}{1995}.

\bibitem{charm2}M.L. Mangano {\it et al.}, \Journal{\NPB}{405}{507}{1993}.

\bibitem{pQCDpi0} B.A. Kniehl {\it et al.},
\Journal{\NPB}{597}{337}{2001}; S. Kretzer,
\Journal{\PRD}{62}{054001}{2000}.

% --=============== shadowing effect
\bibitem{lin96} Z. Lin and M. Gyulassy,
        \Journal{\PRL}{77}{1222}{1996}.

% --=============== charm energy loss
\bibitem{dokshitzer01} Y.L. Dokshitzer and D.E. Kharzeev,
        \Journal{\PLB}{519}{199}{2001}.

% --=========== Loic
\bibitem{loic} L. Grandchamp and R. Rapp, \Journal{\PLB}{523}{60}{2001}.

%--====== PBM
%\bibitem{pbm} A. Andronic, P. Braun-Munzinger, K. Redlich, J. Stachel,
%        \Journal{\PLB}{571}{36}{2003}, and references therein.
\bibitem{pbm} A. Andronic {\it et al.},
        \Journal{\PLB}{571}{36}{2003}.


%--======= Rafleski
\bibitem{rafelski} R.L. Thews, M. Schroedter, and J. Rafelski,
        \Journal{\PRC}{63}{054905}{2001}.

%--======= McLerran
\bibitem{mclerran}  M.I. Gorenstein {\it et al.}, \Journal{\JPG}{28}{2151}{2002}.

%j/psi melting - qgp signal
\bibitem{matsui} T. Matsui and H. Satz,
        \Journal{\PLB}{178}{416}{1986}.

\bibitem{xu2}
%N. Xu and Z. Xu \Journal{\NPA}{715}{587c}{2003};
        S. Batsouli {\it et al.}, \Journal{\PLB}{557}{26}{2003};
    Z.W. Lin and D. Molnar, \Journal{\PRC}{68}{044901}{2003};
    V. Greco, C.M. Ko and R. Rapp, \Journal{\PLB}{595}{202}{2004};
    X. Dong {\it et al.}, \Journal{\PLB}{597}{328}{2004}.

%--========================== review paper of charm
\bibitem{tavernier} S.P.K. Tavernier,
\Journal{\RPP}{50}{1439}{1987}. Total cross sections extrapolated from
very high $x_F$ and/or from correlations with low acceptance are not
included in Fig.\ref{fig4e}.

%--=========================== Fermi Lab Charm
\bibitem{fermic} G.A. Alves {\it et al.} (E769 Collaboration),
        \Journal{\PRL}{77}{2388}{1996}.

%--========================== ISR electron Charm,...
\bibitem{isrc} F.W. B\"{u}sser {\it et al.}, \Journal{\NPB}{113}{189}{1976}.
%(XXX needs Charm-hadron references as well XXX)

%--=========================== CDF II
\bibitem{CDFII} D. Acosta {\it et al.} (CDF II Collaboration),
        \Journal{\PRL}{91}{241804}{2003}.

%---==== phenix electron, 130 GeV Au+Au
\bibitem{phenixe} K. Adcox {\it et al.} (PHENIX Collaboration),
        \Journal{\PRL}{88}{192303}{2002}.

% --==== UA2
\bibitem{ua2} O. Botner {\it et al.}, \Journal{\PLB}{236}{488}{1990}.

% --======== systematic charm at RHIC
\bibitem{vogt02} R. Vogt, \Journal{\IPE}{12}{211}{2003}; R. Vogt, hep-ph/0203151.
%In Fig.~\ref{fig4e}, the solid
%line  depicts a NLO pQCD calculation with MRST HO,
%$\mu_{F}=\mu_{R}=2m_{c}$, $m_{c}$ = 1.2 GeV/$c^{2}$. Other two
%dot-dashed lines depict the calculations with the only change
%$\mu_{R}=m_{c}$ and $\mu_{R}=0.5m_{c}$, respectively.

% --======== dipole model
\bibitem{dipole} J. Raufeisen and J.-C. Peng,
\Journal{\PRD}{67}{054008}{2003}.

% --======== dAu high pt PRL paper
\bibitem{dAuhighpt}J. Adams {\it et al.} (STAR Collaboration), \Journal{\PRL}{91}{072304}{2003}.

% --============= pi, K, pr, TOFr
\bibitem{startof1} J. Adams {\it et al.} (STAR Collaboration),
nucl-ex/0309012.

% --============================= tpc NIM paper
\bibitem{tpcnim} M. Anderson {\it et al.},
  \Journal{\NIMA}{499}{659}{2003}.

% --=============================== Kstar* paper
\bibitem{kstarprc} C. Adler {\it et al.} (STAR Collaboration),
        \Journal{\PRC}{66}{061901(R)}{2002};
        %H. Zhang, \Journal{\JPG}{30}{S577}{2004};
        H. Zhang, Ph.D. thesis, Yale University, 2003.

%
\bibitem{hijing} X.N. Wang and M. Gyulassy,
   \Journal{\PRD}{44}{3501}{1991}.
%

% --=============================== TOF nim paper
\bibitem{tofnim} B. Bonner {\it et al.},
   \Journal{\NIMA}{508}{181}{2003}; M. Shao {\it et al.},
   \Journal{\NIMA}{492}{344}{2002}.

% --=============Photon and neutral pion production
\bibitem{johnson} J. Adams {\it et al.} (STAR Collaboration),
\Journal{\PRC}{70}{044902}{2004}; I.  Johnson, Ph.D. thesis,
U.C. Davis, 2002.

\bibitem{pythia1} T. Sj\"ostrand {\it et al.}, \Journal{Computer
Physics Commun.}{135}{238}{2001}.  PYTHIA 6.152 was used with the
parameter settings: MSEL=1, CTEQ5M1.  Bottom contribution was
estimated to be ~20-30\% in $p_T$=2-3 GeV/$c$, ~40-50\% in 3-4 GeV/$c$
to the total non-photonic electrons, and negligible at lower $p_T$.

%\bibitem{qgp} H. Satz, \Journal{\NPA}{715}{3c}{2003}.

% --================================ PDG
\bibitem{pdg} K. Hagiwara {\it et al.},
\Journal{\PRD}{66}{010001-256}{2002} (Particle Data Group).
$N_{D^{0}}/N_{c\bar{c}}$ was derived from the measured open charm
states in $e^{+}e^{-}$ at $\sqrt{s}=91$ GeV. There is a 10\%
uncertainty taken into account in the branching ratios of open
charm semileptonic decays used in the electron-to-charm fit,
reflecting unknown production and branching rations for individual
states. The ratio in hadronic production, although somewhat
reduced, is consistent with that in $e^{+}e^{-}$ within errors.

%--===== cosmic data
\bibitem{cosmic}I.V. Rakobolskaya {\it et al.}, \Journal{\NPB}{122}{353}{2003}
(Proc. Suppl.).

%-===============
%\bibitem{CDF2} F. Abe {\it et al.},
%\Journal{\PRD}{41}{2330R}{1990}.

%--===== PHENIX Jpsi in AuAu
\bibitem{phenixjpsi} S.S. Adler {\it et al.} (PHENIX Collaboration),
\Journal{\PRC}{69}{014901}{2004}; S.S. Adler {\it et al.} (PHENIX
Collaboration), \Journal{\PRL}{92}{051802}{2004}.

\end{thebibliography}
\end{document}